\documentclass[journal=chreay,manuscript=review,layout=twocolumn]{achemso}

\usepackage[version=3]{mhchem} 
\usepackage{amsmath}
\usepackage{amssymb}
\usepackage{graphicx}
\usepackage{flushend} 

\usepackage{nicefrac}
\setkeys{acs}{maxauthors = 99} 

\usepackage{mathtools}

\DeclarePairedDelimiter\ket{\lvert}{\rangle}
\DeclarePairedDelimiterX\braket[2]{\langle}{\rangle}{#1 \delimsize\vert #2}

\usepackage[colorlinks=true, allcolors=black]{hyperref}
\usepackage{siunitx}
\usepackage{amsfonts}

\title[Electric field lifetime tuning]
  {Electric-field control of photon indistinguishability in cascaded decays in quantum dots}

\author{Gabriel Undeutsch}
\email{gabriel.undeutsch@jku.at}
\affiliation[JKU Linz]
{Institute of Semiconductor and Solid State Physics, Johannes Kepler University, Linz, Austria}
\author{Maximilian Aigner}
\affiliation[JKU Linz]
{Institute of Semiconductor and Solid State Physics, Johannes Kepler University, Linz, Austria}
\author{Ailton J. Garcia Jr.}
\affiliation[JKU Linz]
{Institute of Semiconductor and Solid State Physics, Johannes Kepler University, Linz, Austria}
\author{Johannes Reindl}
\affiliation[JKU Linz]
{Institute of Semiconductor and Solid State Physics, Johannes Kepler University, Linz, Austria}
\author{Melina Peter}
\affiliation[JKU Linz]
{Institute of Semiconductor and Solid State Physics, Johannes Kepler University, Linz, Austria}
\author{Simon Mader}
\affiliation[JKU Linz]
{Institute of Semiconductor and Solid State Physics, Johannes Kepler University, Linz, Austria}
\author{Christian Weidinger}
\affiliation[JKU Linz]
{Institute of Semiconductor and Solid State Physics, Johannes Kepler University, Linz, Austria}
\author{Saimon F. Covre da Silva}
\affiliation[JKU Linz]
{Institute of Semiconductor and Solid State Physics, Johannes Kepler University, Linz, Austria}
\alsoaffiliation[Campinas]
{Instituto de Física Gleb Wataghin, Universidade Estadual de Campinas, Campinas, Brazil}
\author{Santanu Manna}
\affiliation[JKU Linz]
{Institute of Semiconductor and Solid State Physics, Johannes Kepler University, Linz, Austria}
\alsoaffiliation[Delhi]
{Department of Electrical Engineering, Indian Institute of Technology Delhi, India}
\author{Eva Schöll}
\affiliation[JKU Linz]
{Institute of Semiconductor and Solid State Physics, Johannes Kepler University, Linz, Austria}
\alsoaffiliation[JKU2]
{Institute for Integrated Circuits and Quantum Computing, Johannes Kepler University, Linz, Austria}
\author{Armando Rastelli}
\affiliation[JKU Linz]
{Institute of Semiconductor and Solid State Physics, Johannes Kepler University, Linz, Austria}
\email{armando.rastelli@jku.at}

\abbreviations{HOM,QD}

\begin{document}






\begin{center}
\parbox{\textwidth}{
\begin{abstract}
Photon indistinguishability, entanglement, and antibunching are key ingredients in quantum optics and photonics. Decay cascades in quantum emitters offer a simple method to create entangled-photon-pairs with negligible multi-pair generation probability.
However, the degree of indistinguishability of the photons emitted in a cascade is intrinsically limited by the lifetime ratio of the involved transitions.
Here we show that, for the biexciton-exciton cascade in a quantum dot, this ratio can be widely tuned by an applied electric field. Hong-Ou-Mandel interference measurements of two subsequently emitted biexciton photons show that their indistinguishability increases with increasing field, following the theoretically predicted behavior. At the same time, the emission linewidth stays close to the transform-limit, favoring applications relying on the interference among photons emitted by different sources. 

\end{abstract}
}
\end{center}

\noindent\hrulefill 
\vspace{1em} 
\twocolumn[{
    \noindent
}]


In the realm of quantum technologies, many applications require specialized quantum light sources that meet stringent criteria. Among the most sought-after properties is the capability of ``on demand'' generation of simultaneously highly indistinguishable and strongly entangled photon pairs \cite{lloyd2001,Kimble2008, Lu2014}.
Epitaxial semiconductor quantum dots (QDs) have emerged as promising candidates for generating photons with high brightness \cite{liu2019, tomm2021, ding2023}, high single-photon purity\cite{schweickert2018}, narrow linewidth\cite{kuhlmann2015,DaSilva2021,zhai2020, laferriere2023}, and near-unity indistinguishability\cite{Huber2015, ding2016, Scholl2019, tomm2021, Zhai2022}.
Additionally, the biexciton (XX) - exciton (X) radiative cascade allows the direct generation of on-demand polarization-entangled photon pairs with near-unity time-averaged fidelities \cite{Muller2014, Jons2017, Huber2018, Xiangjun2021, Schimpf2023, chen2024}.
However, the cascade nature of the process to create entangled photon pairs causes an unwanted temporal entanglement between the two photons, resulting in a non-separable two-photon state. This reduces the state purity $\mathbb{P}$, describing the indistinguishability of the emitted single photons in the time-domain, to \cite{simon2005, Huber2013, Scholl2020} 
\begin{equation}
    \mathbb{P} =  \frac{1}{r+1},
    \label{equ:Scholl}
\end{equation}
with the ratio of the radiative lifetimes ${r=\tau_{XX}/\tau_{X}}$. 
The purity is experimentally not directly accessible, but -- for systems with negligible multi-photon probability, as is the case here -- it is identical to the two-photon interference visibility\cite{Fischer2018}, which can be measured in Hong-Ou-Mandel (HOM) type experiments.
The observed QD lifetime ratio is typically $r\approx 0.4-0.7$\cite{Muller2014, Schimpf2019, DaSilva2021, Scholl2020}, resulting in a maximum achievable HOM interference visibility of $0.67$.
It has been proposed and demonstrated that a suitable optical cavity can selectively shorten the XX state ($\ket{XX}$) lifetime while keeping the X state ($\ket{X}$) lifetime constant, and thereby decrease the lifetime ratio\cite{Huber2013}. However, no increase in photon indistinguishability has yet been shown. 
In this work, we take a different approach and demonstrate that the lifetime ratio $r$ can be conveniently modified by applying a vertical electric field to QDs embedded in a p-i-n diode.
The diode structure allows for the charge control of the QD and its environment, enabling stabilization and tuning of the emission properties~\cite{Warburton2000, patel2010, kaniber2011, bennett2010a, kuhlmann2015,zhai2020,Zhai2022, Schimpf2021}. In particular, it has been demonstrated that an electric field induces a non-monotonic variation in the $\ket{X}$ lifetime \cite{polland1985,fry2000}, but we are not aware of similar measurements for the $\ket{XX}$.

We focus on GaAs QDs obtained by local droplet etching epitaxy \cite{Heyn2009,DaSilva2021}, as these QDs have recently outperformed other systems in terms of degree of polarization-entanglement~\cite{Keil2017,Schimpf2021,Huber2018}, single photon purity~\cite{schweickert2018}, photon indistinguishability~\cite{Scholl2019,zhai2020,Zhai2022}, and spin properties~\cite{Appel2025}. 
We find that an increasing electric field leads to a monotonically increasing $\ket{X}$ lifetime, while the $\ket{XX}$ lifetime remains almost unchanged. This allows us for the first time to tune the lifetime ratio by an externally applied electric field. With that, we decrease the ratio from 0.64(2) to 0.25(2) and thus increase the theoretically maximum indistinguishability from 0.61(1) to 0.794(8), which is significantly higher than the value typically achieved in non-diode structures.
We experimentally verify this by measuring the HOM visibility for XX and X photons. At high fields, the HOM visibility for X photons degrades, most likely due to residual charge noise in the sample, as indicated by linewidth broadening. However, the XX visibility closely follows the theoretical prediction, with its linewidth remaining close to the transform-limit even at high fields.

\begin{figure*}[htb]
\centering
\includegraphics[width=\textwidth]{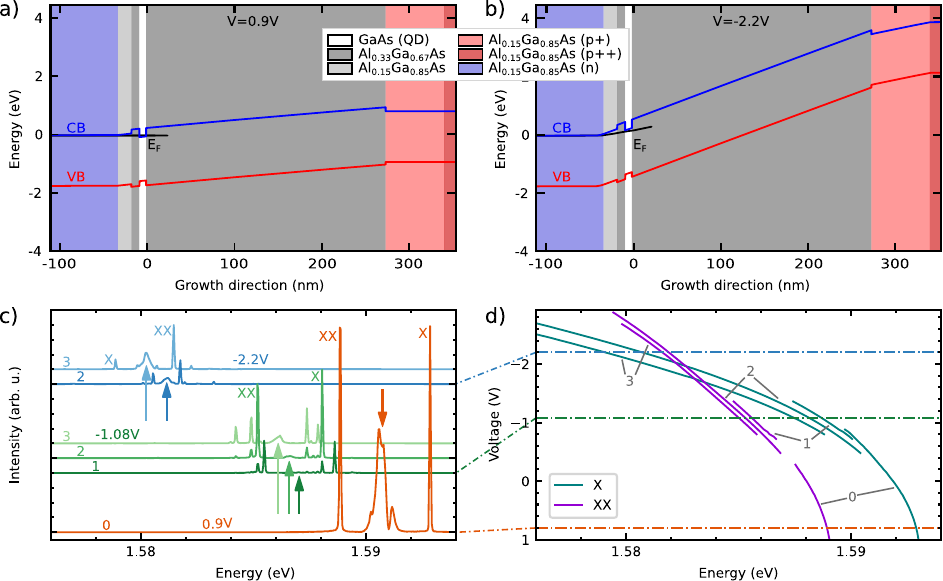}
\caption{Sample structure with simulated conduction band (CB) and valence band (VB) edges and CB quasi Fermi level (E$_F$) for a) an applied voltage $V=\SI{0.9}{V}$ and b) $V=\SI{-2.2}{V}$. c) Spectra at three different voltages under $\pi$-pulse two-photon excitation (TPE). Depending on the voltage, several biexciton-exciton (XX-X) replicas (labeled 0-3) could be observed, sometimes simultaneously. This is most likely due to different numbers of holes trapped in the vicinity of the QD. For $V=\SI{0.9}{V}$ (orange) only replica 0 is visible. For $V=\SI{-1.08}{V}$ (green) and $V=\SI{-2.2}{V}$ (blue), three replicas (1--3) and two replicas (2 and 3) are visible. For each spectrum, the laser energy (indicated with arrows) was adjusted to match the TPE resonance. Small contributions from other replicas come from phonon-assisted excitation. d) Fitted energy of the XX and X photons as a function of voltage. Each data point is extracted from a spectrum under TPE of the respective replica.}
\label{fig: Basics}
\end{figure*}

The GaAs QDs used in this work are embedded into a p-i-n diode structure (see Fig.~\ref{fig: Basics}~a) and b)) and a weak planar cavity built up of distributed Bragg reflectors, with ten pairs below and four pairs above the QD layer (see supplementary). 
When an external voltage $V$ is applied to the diode structure, an electric field $F_V\simeq \frac{V_b-V}{D}$ is generated, where $V_b$ is the built-in voltage and $D$ is the thickness of the intrinsic layer \cite{fry2000}. (Note that the n-doped layer is grounded and $V$ is the voltage applied to the top p-doped layer.) In our case, we expect $V_b$ to be about +\SI{1.7}{V} as the Al$_{15}$Ga$_{85}$As band gap energy at low temperatures is about \SI{1.73}{eV}. A higher electric field (lower voltage) leads to a stronger bending of the conduction (CB) and valence band (VB) edges, as seen from the comparison between the calculation results shown in Fig.~\ref{fig: Basics}~a) and b) for $V=\SI{0.9}{V}$ and $V=\SI{-2.2}{V}$, respectively. Since in the latter case the energy $E_F$ of the CB quasi-Fermi level lies below the CB edge in the QD region, the QD is in a neutral state and theoretically only neutral excitonic states can be excited.

To characterize the sample, we perform voltage-dependent photoluminescence measurements of a single QD (QD~1) under resonant excitation of $\ket{XX}$ via two-photon excitation (TPE). 
Representative spectra, collected at different voltages and excitation energies are shown in Fig.~\ref{fig: Basics}~c). For positive voltages, we observe the typical spectrum of GaAs QDs (see example in orange for $V=+\SI{0.9}{V}$), with the dominant XX and X lines stemming from the radiative cascade. 
At negative voltages, we find several XX and X replicas with slightly different emission energies, see green and blue spectra in Fig.~\ref{fig: Basics}~c). We attribute such replicas to variations in the electric field~\cite{Houel2012}, caused by different numbers of holes caught at the tunnel barrier interface close to the QD layer (see supplementary). We study the three most prominent XX-X replicas (labeled as 1--3 according to their energy), which we address by tuning the laser energy to resonantly excite the respective $\ket{XX}$. Additional small lines come from other cascades due to phonon-assisted TPE. 
Figure~\ref{fig: Basics}~d) shows the fitted emission energies for the XX and X photons for varying voltage (similar data for another QD is shown in the supplementary).
In a first approximation, the field dependence of the emission energy can be described by a quadratic behavior\cite{fry2000}, similar to the potential energy of a polarizable electric dipole in an external electric field (see supplementary). Most importantly, we see that the X line red-shifts faster than the XX line with increasing electric field (decreasing voltage). This observation, which is consistent with previous results on InGaAs QDs~\cite{Trotta2013, patel2010}, indicates that the $\ket{X}$ can be more easily polarized than the $\ket{XX}$. Intuitively, we attribute this observation to the larger number of charge carriers present in the biexciton complex, partly screening the external field. As a result, for sufficiently large negative voltages, the XX and X emission lines cross and swap their order, as illustrated by the spectra of replicas 2 and 3 at $V=\SI{-2.2}{V}$ in Fig.~\ref{fig: Basics}~c).

\begin{figure*}[htb]
\centering
\includegraphics[width = \textwidth]{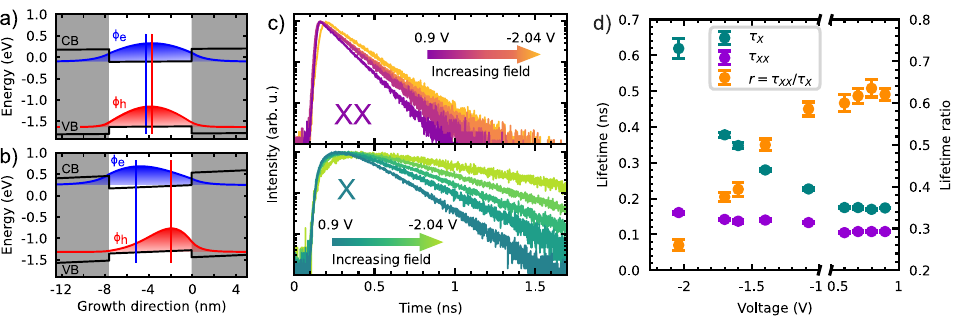}
\caption{Schematic of electron $\phi_e$ and hole $\phi_h$ wave functions, showing a) a high overlap under small electric field vs. b) a reduced overlap under high electric field. c) Lifetime measurements of the $\ket{XX}$ and $\ket{X}$ for increasing electric fields (indicated by an arrow) in the specified voltage range. d) Fitted lifetimes $\tau$ as a function of the gate voltage. For decreasing voltage, the relative increase of $\tau_{XX}$ is much lower than that of $\tau_{X}$, resulting in a decrease of the lifetime ratio.}
\label{fig: Lifetimes}
\end{figure*}

In addition to charge and energy tuning, the electric field influences the overlap of the electron and hole wave functions $\braket{\Phi_e}{\Phi_h}$. From a quadratic fit of the X energy we find that, at an applied voltage of $V\approx+\SI{1.1}{V}$, the permanent dipole~\cite{fry2000, barker2000,jin2004, Finley2004, mar2017} present at zero field cancels with the induced dipole, leading to a near maximum achievable wave function overlap, as shown by the schematic in Fig.~\ref{fig: Lifetimes}~a). (Note that in the studied device we cannot reach this point, since for $V\gtrsim\SI{1}{V}$ X and XX luminescence is quenched due to single electron charging.) 
Any change in voltage from this point will pull the wave functions in opposite directions, resulting in a reduced overlap, as sketched in Fig.~\ref{fig: Lifetimes}~b).
In a single particle picture, we expect the decay rate to be proportional to the overlap integral of the electron and hole wave functions \cite{Bastard1983}. The change in overlap in response to a change in electric field, in turn, depends on the polarizability of the excitonic species. From the observation that the XX line shifts less than the X line for increasing electric field, we can already anticipate that the $\ket{X}$ lifetime will increase more than the $\ket{XX}$ lifetime with increasing field. 
This expectation is confirmed by measuring the dynamics of the XX and X emission following TPE, as shown in Figs.~\ref{fig: Lifetimes}~c). Here and in the following measurements, we always use the brightest replica at a given voltage (see supplementary). Figure~\ref{fig: Lifetimes}~d) shows the lifetimes extracted from a fit of the data (see supplementary) as a function of the applied voltage, as well as the resulting lifetime ratio $r$. For positive voltages, both lifetimes stay almost constant with $\tau_{XX}\approx\SI{110}{ps}$ and $\tau_{X}\approx\SI{175}{ps}$. For negative voltages, the $\ket{XX}$ lifetime increases by a factor of 1.5 to $\SI{161(4)}{ps}$, while the lifetime of the $\ket{X}$ increases significantly by a factor of 3.5 to $\SI{619(27)}{ps}$ at $V=\SI{-2.04}{V}$. 
Consequently, the lifetime ratio decreases from 0.64(2) to 0.26(1), as shown in orange in Fig.~\ref{fig: Lifetimes}~d). Thus, from Eq.~(\ref{equ:Scholl}), the theoretical limit for the indistinguishability increases.

\begin{figure*}[htb]
\centering
\includegraphics[width=\textwidth]{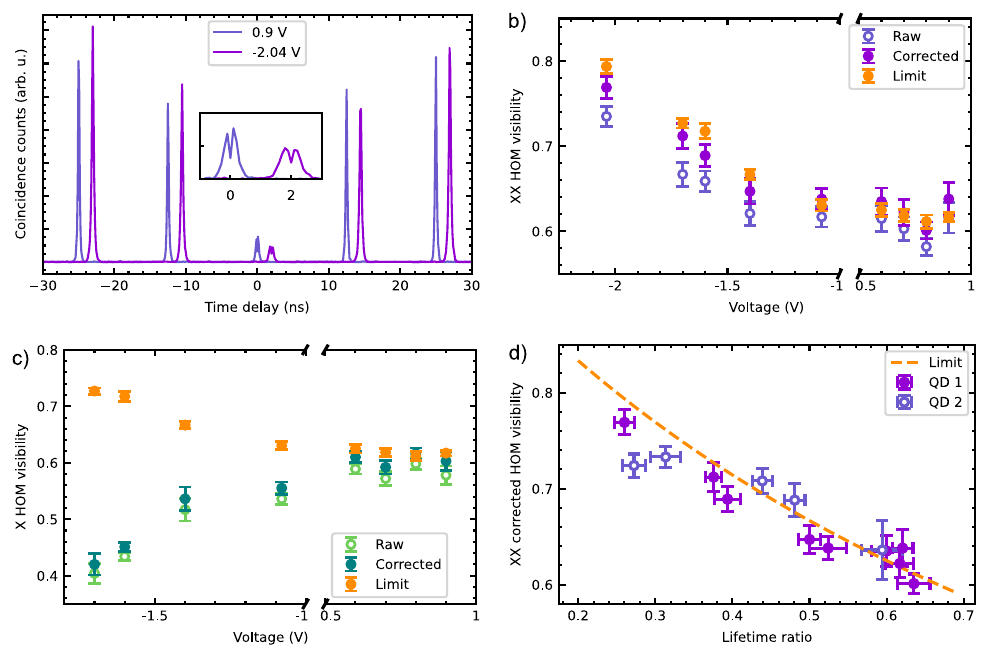}
\caption{a) Histograms of Hong-Ou-Mandel (HOM) measurements of the XX photons for the two indicated voltages, with one shifted by \SI{2}{ns} for better readability. Raw and corrected HOM visibilities for b) the XX and c) the X for different applied voltages, as well as the theoretical limit given by Eq.~(\ref{equ:Scholl}). Whereas the XX follows the expected trend, the X HOM visibility decreases with voltage. d) Corrected HOM visibility of the XX as a function of lifetime ratio $r=\tau_{XX}/\tau_{X}$.  In addition to QD~1, used for all measurements in the manuscript, we also show data obtained from another QD (QD~2). }
\label{fig: HOM}
\end{figure*}

To experimentally investigate whether predictions are correct, we measure the two-photon interference visibility for two sequentially emitted XX or X photons in a HOM type interferometer (see supplementary for measurement and analysis details) with a time delay matching the repetition rate of the excitation laser and for different applied voltages. To benchmark the setup and the QDs, we measure the HOM visibility of the resonantly excited negative trion from the same QD at a gate voltage of $V=+\SI{1.03}{V}$, since this is not intrinsically limited by a cascaded emission. From such a measurement, a raw visibility of 0.944(4) and a corrected visibility of 0.991(6) are obtained (see supplementary).
Representative photon coincidence histograms for the XX line at $V=+\SI{0.9}{V}$ and $\SI{-2.04}{V}$ are shown in Fig.~\ref{fig: HOM}~a) with a horizontal shift of \SI{2}{ns} for better readability. A decreased central peak is clearly visible for $V=\SI{-2.04}{V}$ in the inset, indicating an improved HOM visibility. The evaluated HOM visibilities for different voltages are shown in Fig.~\ref{fig: HOM}~b) and d) for the XX and in Fig.~\ref{fig: HOM}~c) for the X together with the theoretical limit for the indistinguishability from Eq.~(\ref{equ:Scholl}).
For positive voltages, both the XX and the X show HOM visibilities of $\approx 0.6$, consistent with the constant lifetime ratio. For negative voltages, the XX follows the expected trend and almost reaches the theoretical limit. The highest measured raw (corrected) HOM visibility is 0.735(12) (0.769(13)) at $V=\SI{-2.04}{V}$ ($r=0.26$), which is close to the theoretical limit of 0.794(8). Figure~\ref{fig: HOM}~d) shows the corrected HOM visibility of the XX against the lifetime ratio. Additionally, it also includes data from a second QD, further supporting our observations.
In contrast to the results obtained for the XX and to the theoretical expectations, the raw (corrected) HOM visibility for the X photons degrades to 0.404(18) (0.420(19)) with decreasing lifetime ratio. 

\begin{figure*}[htb]
\centering
\includegraphics[width=\textwidth]{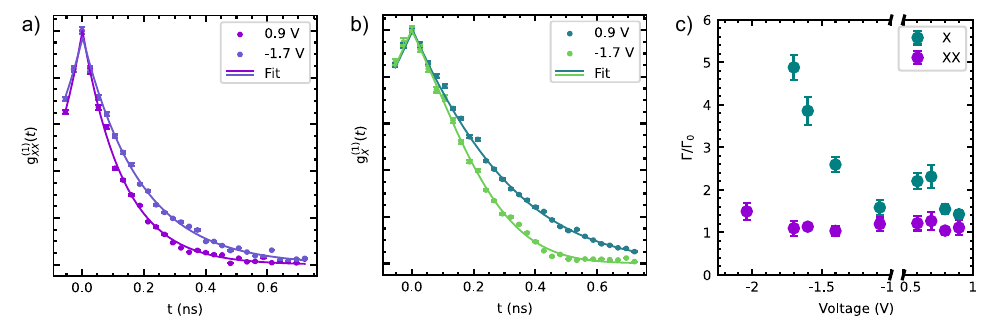}
\caption{First-order correlation function $g^{(1)}(t)$ as a function of the relative delay $t$ of a) XX and b) X for two different applied voltages, recorded using a Michelson interferometer. c) Fitted linewidths relative to the respective Fourier transform limit as function of voltage, showing almost no line broadening for the XX but significant broadening for the X.}
\label{fig: Michelson}
\end{figure*}

To understand the origin of the indistinguishability drop for X photons and provide a more stringent measurement of the optical quality of the QD emitter  at different electric fields, we measure the coherence time of the XX and X lines using a Michelson interferometer (see supplementary). In absence of noise, the upper limit of the coherence time for an X photon is given by $\tau_c=2\tau_{X}$, leading to a transform-limited linewidth of $\Gamma_{0,X}=\frac{\hbar}{\tau_{X}}$. For the XX, we expect instead $\Gamma_{0,XX}=\hbar\left(\frac{1}{\tau_{X}}+\frac{1}{\tau_{XX}}\right)$ \cite{Chiang2010}.
Figure~\ref{fig: Michelson}~a) and b) show the first-order correlation function $g^{(1)}(t)$ for the XX and the X respectively for two voltages. At a gate voltage of $V=+\SI{0.9}{V}$, the transform-limited linewidths for the XX (X) are $\SI{9.88(12)}{\mu eV}$ ($\SI{3.78(4)}{\mu eV}$). The linewidths from the Michelson measurements are $\SI{11.0(1.7)}{\mu eV}$ ($\SI{5.4(4)}{\mu eV}$). Therefore, the XX (X) transition is only a factor $\Gamma/\Gamma_0=1.1(2) (1.4(1))$ away from the transform-limit.
At a gate voltage of $V=\SI{-1.7}{V}$, the measured linewidth of the XX is $\SI{7.0(1.1)}{\mu eV}$. Together with the transform-limited linewidth $\Gamma_0=\SI{6.38(10)}{\mu eV}$, this yields the same factor of 1.1(2) as for positive voltages. For the X, the measured linewidth increases to $\SI{8.5(5)}{\mu eV}$, whereas $\Gamma_0$ decreases to $\SI{1.74(3)}{\mu eV}$. The X linewidth is therefore a factor 4.9(3) away from the transform-limit. Figure~\ref{fig: Michelson}~c) shows the ratio $\Gamma/\Gamma_0$ for different gate voltages. It is interesting to note that the XX linewidth stays close to the transform-limit over the whole voltage range, suggesting that photon indistinguishability is preserved over time separations extending to several minutes (the typical duration of a Michelson interferometry measurement). 
In contrast, the X linewidth broadens significantly for decreasing voltages, in line with the drop in HOM indistinguishability shown in Fig.~\ref{fig: HOM}~b). We attribute these observations to residual charge noise and the higher sensitivity of the X transition energy to noise because of its higher polarizability compared to the XX transition energy~\cite{Schimpf2019}. 

In summary, we have demonstrated that the lifetime of transitions in the decay cascade of QDs can be differentially tuned using an external electric field. This reduces the intrinsic limitations on indistinguishability, as confirmed experimentally for the XX photons emitted by GaAs QDs.
By operating a p-i-n diode with embedded QDs at negative voltages, strong band bending reduces the overlap of electron and hole wave functions, leading to an increased excited state lifetime. This effect is more pronounced for the $\ket{X}$ compared to the $\ket{XX}$, due to the higher sensitivity of the exciton to electric field changes compared to the biexciton complex. Our measurements show a reduction in the lifetime ratio from 0.64(2) to 0.26(1). This results in an improved (corrected) HOM interference visibility of 0.769(13) for the XX, approaching the theoretical limit of 0.794(8) -- well beyond the values achievable in absence of an electric field. However, the X HOM visibility decreases as the lifetime ratio decreases, which we attribute to an increased sensitivity to noise. 
Achieving a degree of indistinguishability well above 0.9 remains essential for quantum technology applications. The tuning range of the lifetime ratio could be further increased by dedicated design of the diode structure.
Additionally, combining a diode structure with a tailored microcavity, can selectively shorten the $\ket{XX}$ lifetime through Purcell enhancement while maintaining the $\ket{X}$ lifetime relatively unchanged.
These findings, which we expect to apply also to other material systems, may contribute to obtain a quantum light source that simultaneously combines the emission of highly indistinguishable photons with low multi-photon probability, transform-limited linewidths and high degree of polarization-entanglement -- all key requirements for advancing quantum networks and other quantum technology applications that have long been anticipated.

\begin{acknowledgement}
We thank Petr Klenovský and Michał Gawełczyk for fruitful discussions.
This project has received funding from the European Union’s Horizon 2020 research and innovation program under Grant Agreement No. 871130 (Ascent+) and the EU HE EIC Pathfinder challenges action under grant agreement No. 101115575, from the QuantERA II program that has received funding from the European Union’s Horizon 2020 research and innovation program under Grant Agreement No. 101017733 via the projects QD-E-QKD and MEEDGARD (FFG Grants No. 891366 and 906046) the Austrian Science Fund FWF via the Research Group FG5, I 4320, I 4380, from the Austrian Science Fund FWF 42through [F7113] (BeyondC), and from the cluster of excellence quantA [10.55776/COE1] as well as the Linz Institute of Technology (LIT), the LIT Secure and Correct Systems Lab, supported by the State of Upper Austria.
S.F.C. da Silva acknowledges São Paulo Research Foundation (FAPESP), Brasil, Process Number 2024/08527-2 for financial support.

\end{acknowledgement}

\begin{suppinfo}
Sample: Sample structure, Band structure simulations; Methods: Cryogenic micro-photoluminescence setup, Autocorrelation measurements, Lifetime measurement, Hong-Ou-Mandel measurements, Michelson interferometry;
Quantum confined Stark effect: Polarizability, Wave function overlap, Permanent dipole;
Replicas of the XX-X paors: Origin of the replicas, Photogeneration of the holes near QDs, Localization of the holes close to the QD, Replica intensities, Comparison of replicas at the same voltage; Additional Data: Indistinguishability of the resonantly excited negative trion, Single-photon purity, Data on additional quantum dots.

\end{suppinfo}

\bibliography{references}
\end{document}


\section{Sample}
\subsection{Sample structure}
\begin{figure}
    \centering
    \includegraphics[width = 0.7\textwidth]{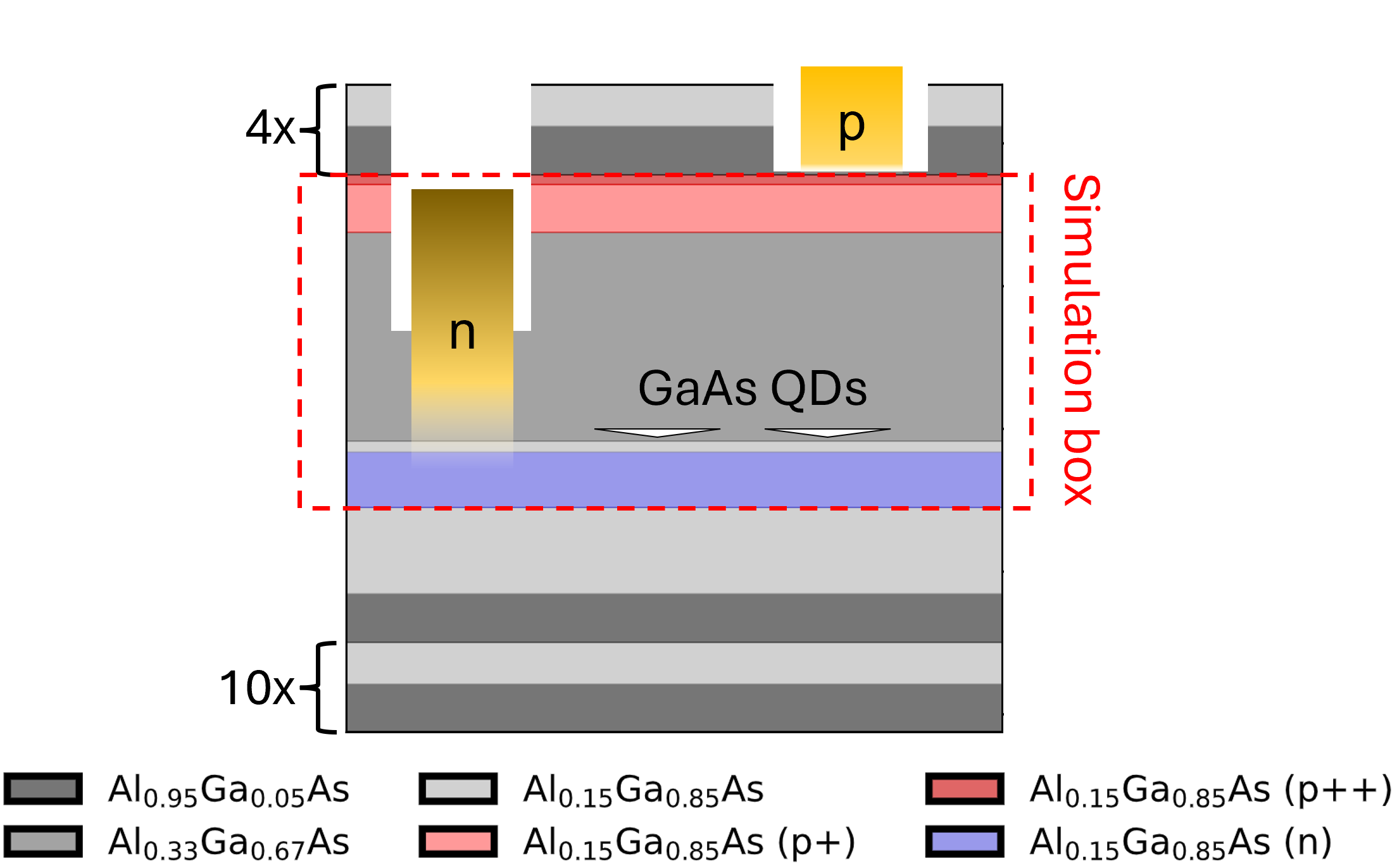}
    \caption{Schematics of the sample structure, including the simulation box.}
    \label{sfig:sample_structure_simulation_box}
\end{figure}

The GaAs quantum dots (QDs) studied in this paper were fabricated using molecular beam epitaxy with local droplet etching (LDE). The schematic sample structure is shown in Fig.~\ref{sfig:sample_structure_simulation_box}. First, a semi-insulating GaAs (001) substrate undergoes in-situ deoxidation, followed by the deposition of a GaAs buffer layer and the growth of a superlattice (30$\times$ (\SI{2.5}{nm} AlAs/\SI{2.5}{nm} GaAs)) to planarize the surface and bury any surface impurities. 
A second GaAs buffer layer and a planar distributed Bragg reflector (DBR) composed of 10 pairs of Al$_{0.95}$Ga$_{0.05}$As/Al$_{0.15}$Ga$_{0.85}$As are then grown. An additional Al$_{0.95}$Ga$_{0.05}$As layer is incorporated to match the QD position with the electric field antinode of the planar 5$\lambda$/2 cavity, which is formed by four additional DBR pairs to improve the extraction efficiency.
The QDs are embedded in a p-i-n diode structure. On the n-side, below the QDs, the diode features a \SI{78}{nm} thick Al$_{0.15}$Ga$_{0.85}$As layer doped with Si ($n_n = 7.4 \cdot 10^{17}~\mathrm{cm}^{-3}$), serving as an electron reservoir for the QDs. The lower 15\% Al concentration prevents the formation of deep donor levels (DX centers) \cite{Mooney1990, Munoz1993}. A tunneling barrier between the n-doped layer and the QDs is formed by a \SI{5}{nm} Al$_{0.15}$Ga$_{0.85}$As layer grown at low temperature to reduce Si segregation, followed by a \SI{10}{nm} Al$_{0.15}$Ga$_{0.85}$As layer grown at higher temperature, as well as a \SI{15}{nm} Al$_{0.33}$Ga$_{0.67}$As layer.
For the QD growth, Al is evaporated onto the Al$_{0.33}$Ga$_{0.67}$As layer without an As background, forming droplets that etch approximately \SI{8}{nm} deep nanoholes in the surface 
during exposure to a reduced As flux. 
These nanoholes are subsequently filled with GaAs to form the QDs, which are then capped with an Al$_{0.33}$Ga$_{0.67}$As layer. Further details are provided in \cite{DaSilva2021}. The QDs are positioned about \SI{275}{nm} below the carbon-doped p+-type layer of the diode, which consists of a \SI{65.6}{nm} thick Al$_{0.15}$Ga$_{0.85}$As layer with low doping concentration  ($n_{p+} = 5 \cdot 10^{18} ~\mathrm{cm}^{-3}$), followed by a \SI{13}{nm} thick highly doped Al$_{0.15}$Ga$_{0.85}$As layer ($n_{p++} = 9 \cdot 10^{18}~\mathrm{cm}^{-3}$). The top DBR is capped with \SI{4}{nm} of GaAs to protect the AlGaAs layers from oxidation.
To contact the n-layer, trenches are etched using sulfuric acid, followed by deposition of Ni/AuGe/Ni/Au and an annealing step at \SI{420}{\celsius} for two minutes in a forming gas environment. For the p-contacts, Pt/Ti/Pt/Au is used, annealed at \SI{400}{\celsius} for two minutes.

\subsection{Band structure simulation}
The simulation results in this work were obtained with nextnano++, a software for semiconductor nanodevices for the self-consistent solution of the Schrödinger, current, and Poisson equation. \cite{trellakis2006, birner2007} For the simulation, the temperature was set to \SI{5}{K}. Fig.~\ref{sfig:sample_structure_simulation_box} shows an illustration of the contacted diode, including the region over which the simulation was performed (simulation box).
\paragraph{Contacts:}
The layer thicknesses and doping concentrations defined in the simulation were taken from the growth protocol. For the intrinsic layer, a background carbon-doping of $1 \cdot 10^{15}~\mathrm{cm}^{-3}$ was assumed to account for the carbon background in the growth chamber. For the donor and acceptor energy, realistic values of \SI{5.8}{meV} and \SI{27}{meV} were used. \cite{donor_silicon, acceptor_carbon} As a result, the free charge carriers in the structure freeze out when running simulations at low temperatures. 

The n- and p-type contacts were defined as Fermi contact, which results in a pinning of the electron and hole quasi Fermi level to the bias applied to the contact. At voltages well below the built-in voltage, the electric field within the intrinsic region is high enough to saturate the electron and hole drift velocity. Therefore, a high field mobility model was used to limit the electron and hole drift velocity to $0.5\cdot10^7~\mathrm{cm/s}$ and, $0.7\cdot10^7~\mathrm{cm/s}$ respectively. \cite{electron_drift_velocity, hole_drift_velocity} 

\paragraph{Energy levels and wave functions:}
By solving the Schrödinger equation numerically, nextnano++ provides solutions for the energy levels and wave functions within a pre-defined quantum region. At the boundaries of the quantum region, Dirichlet boundary conditions were applied. 
In the one-dimensional simulation, the GaAs QD was included by a quantum well with a thickness of \SI{7.5}{nm}. The thickness was chosen such that the calculated transition energy between the lowest electron and heavy hole state roughly matches the measured exciton (X) emission energy. 

\paragraph{Observations:}
The simulated Stark shift of the quantum well is a symmetric function of the electric field, i.e. the emission energy reaches its maximum at zero field, when the applied voltage is equal to the built-in voltage of the diode ($\approx\SI{1.7}{V}$). In reality, we observe an asymmetric Stark shift for the quantum dots in our diode structures, i.e. the maximum emission energy is reached at voltages well below the built-in voltage. We attribute this behavior to the existence of a permanent dipole moment of the charge complexes in the quantum dot.

For negative voltages, the intrinsic region is fully depleted, and the bands are strongly tilted. For positive voltages, the bands become flatter and the charge carrier density in the intrinsic region increases. When the electron density is sufficiently high in the vicinity of the QD, the electron quasi Fermi level is stable in this region and becomes insensitive to small voltage changes. 
At negative voltages, the electron and heavy hole wave functions are pulled to opposite sides of the quantum well, leading to a reduced overlap.

\section{Methods}
\subsection{Cryogenic micro-photoluminescence setup}
\begin{figure}
    \centering
    \includegraphics[width = \textwidth]{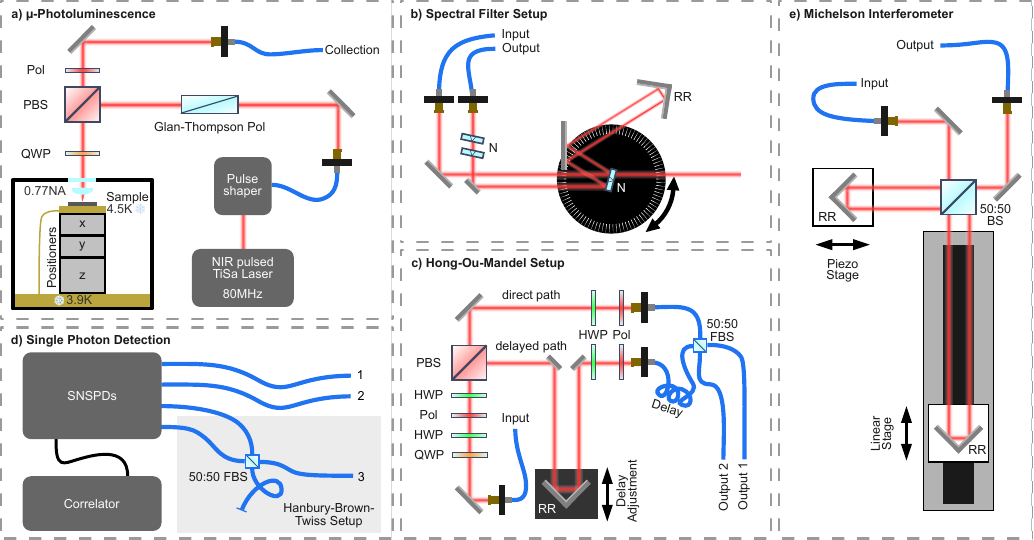}
    \caption{Sketch of the used setup. The parts are separated as indicated and are connected during the measurements using optical fibers. a) cryogenic micro Photoluminescence setup b) spectral filter setup based on volumetric Bragg grating notch filters (N), c) Hong-Ou-Mandel setup to measure the two-photon interference. d) Single photon detection with superconducting nanowire single-photon detectors (SNSPD) and correlator, as well as the Hanbury Brown and Twiss setup to measure the single-photon purity. e) Michelson interferometer to measure the coherence length of the photons. Pol: (linear) polarizer, PBS: polarizing beam splitter, Q(H)WP: quarter (half) wave plate, RR: retro reflector, (F)BS: (fiber) beam splitter}
    \label{sfig:setup}
\end{figure}
Our cryogenic micro photoluminescence spectroscopy setup is shown in Fig.~\ref{sfig:setup}~a). The sample is cooled to \SI{4.5}{K} in a dry closed-cycle cryostat and can be moved in x, y and z directions using nano positioners. 
The QD is excited using a pulsed titanium-sapphire laser with a repetition rate of \SI{80}{MHz}, which is shaped to a pulse length of about \SI{5}{ps} using a 4f pulse shaper. The laser is sent through a Glan-Thompson polarizer to set a well-defined s-polarized state, which is then reflected at the subsequent polarizing beam splitter cube (PBS). This directs the beam into the cryostat, where it is focused onto the sample using a NA=0.77 cold lens. 
The same lens is used in our setup with confocal geometry to collect the emission and back reflected laser. The laser is filtered by cross-polarization, and only the p-polarized emission of the sample passes the PBS and an additional linear polarizer (Pol). The detected signal is coupled into a single mode (SM) fiber and can be flexibly sent to different filtering and analysis setups.
A quarter wave plate (QWP) between the PBS and cryostat compensates small birefringence effects to improve the laser suppression. 
The spectra are recorded using a CCD camera connected to a spectrometer (not shown).
For all correlation measurements described below, spectral filtering of the single photons is necessary. We use volumetric Bragg grating notch filters (N), which reflect a narrow spectral range of about \SI{200}{pm}, while transmitting the rest. 
The central wavelength of reflection can be adjusted by changing the incidence angle and can be tuned by roughly \SI{15}{nm}. The setup shown in Fig.~\ref{sfig:setup}~b) features one N on a rotation stage reflecting the desired wavelength and two to four further N for additional filtering of unwanted spectral components. This filtering setup has a fiber-to-fiber efficiency of $\approx 50\%$.

\subsection{Autocorrelation measurements}
\label{subsec:autocorrelation}
The measurement of the second-order autocorrelation function $g^{(2)}(t)$ is realized in a Hanbury Brown and Twiss experiment.
After spectral filtering, the QD emission is sent into a fiber-coupled 50:50 beam splitter (BS) (input 3 in Fig.~\ref{sfig:setup}~d)). The BS outputs are connected to two superconducting nanowire single photon detectors (SNSPDs). The detection events are recorded as time tags by a time tagger, which are then correlated using a PC, to build a histogram of coincidences as a function of time delay between the events. The detectors have time jitters of 13-\SI{14}{ps}. Together with the time jitter of the time tagger, this results in a total jitter of about \SI{15}{ps} for single events and about \SI{21}{ps} for correlations between two detectors.
The single-photon purity can be extracted by evaluating the number of coincidences in the central peak at $t=0$ relative to the average number of coincidences in the 5 side peaks on both sides. 

\subsection{Lifetime measurements}
Lifetime measurements of the exciton and biexciton states ($\ket{X}$, $\ket{XX}$) are performed by time-correlated single photon counting.
The filtered QD emission is sent to one SNSPD. A second output of the pulse shaper setup is used to tap off the roughly $\SI{5}{ps}$ long excitation laser pulses and send it to another SNSPD. The arrival time of the QD photons with respect to the excitation laser pulse is extracted by correlating the time tags of both detectors. The finite rise time of the laser and finite setup timing resolution are taken into account by measuring the instrument response function (IRF). 
For the XX decay, we assume a monoexponential decay with the $\ket{XX}$ lifetime $\tau_{XX}$. The X decay follows a biexponential decay, with the $\ket{XX}$ and $\ket{X}$ lifetimes as time constants. Both formulas are convoluted with the IRF to fit the obtained histograms.

\subsection{Hong-Ou-Mandel measurements}
According to the Hong-Ou-Mandel (HOM) effect, if two indistinguishable photons simultaneously impinge on the two different entry ports of a 50:50 beam splitter (BS), they will interfere and will always exit the BS together at the same output port. The measurement of the HOM visibility is therefore a good measure for the photon indistinguishability.

\subsubsection{Setup}
The HOM interference visibility of subsequent photons from one source is usually measured using an unbalanced Mach-Zehnder-interferometer, as shown in Fig.~\ref{sfig:setup}~c).
The stream of filtered single photons is sent into the setup using a SM fiber (Input). 
To correct for the random unitary transformation of the polarization by the fiber, a combination of half-wave plate (HWP) and quarter-wave plate (QWP) is used to maximize the transmission through the following polarizer (Pol). 
The HWP in front of the PBS is set to send 50\% of the intensity in each interferometer arm. The arms have different lengths to introduce a delay of
\SI{12.5}{ns} in one arm, which matches the \SI{80}{MHz} repetition rate of the excitation laser. This delay is achieved by a combination of a \SI{2}{m}-long fiber and an adjustable free-space delay. Each path contains a HWP and Pol before a polarization maintaining (PM) fiber connected to a fiber-coupled 50:50 BS, where the interference will take place.
The Pols set the polarization of the photons entering the BS and are set to the same (perpendicular) polarization for co (cross)-polarized measurements. Furthermore, they are aligned along one polarization axis of the fiber. 
The HWPs compensate for the different efficiencies in the two arms and are adjusted such that the total transmitted intensity from each arm is equal. 
The two exits (Output 1 and Output 2) are connected to two SNSPDs (1 and 2 in Fig.~\ref{sfig:setup}~d)) and correlated as described in Sec.~\ref{subsec:autocorrelation}.

\subsubsection{Measurements and data evaluation}
To determine the HOM visibility, correlation measurements with co- and cross-polarization have to be carried out.
In the case of the co-polarized measurement, the photons are ideally indistinguishable, interfere, and thus always leave the BS together. This leads to no coincidences at zero time delay after correlation of the two outputs. For the cross-polarized measurement, the photons are fully distinguishable, and therefore leave the BS at opposite outputs with a probability of 50\%. The expected relative intensities of the central histogram peaks are therefore ...~1,~1,~0.75,~0(0.5),~0.75,~1,~1~... for fully indistinguishable (distinguishable) photons.

To analyze the data, first, the center peak area $A_{\text{center, } j}$ for each measurement ($j \in \{\text{co, cross}\}$) is normalized with the average side peak area for absolute delays $\geq 2\cdot\SI{12.5}{ns}$: $a_j=\frac{A_{\text{center, }j}}{A_{\text{side, }j}}$. The raw HOM visibility can then be determined by 
\begin{equation}
    V_{\text{HOM,raw}} = 1-\frac{a_{\text{co}}}{a_{\text{cross}}}\label{sequ:rawvis}.
\end{equation}
The analysis described here is only valid for a perfect interference BS with splitting ratio $R:T = 50:50$, perfect classical interference visibility $\nu=1$, and perfect single photons ($g^{(2)}(0)=0$). For deviations, the obtained raw HOM visibility can be corrected by\cite{santori2002, Zhai2022}:
\begin{equation}
V_{\text{HOM,corr}}~=~V_{\text{HOM,raw}}\frac{1+2g^{(2)}(0)}{\nu^2}.
\end{equation}
By sending a narrow cw laser through our setup we measured a classical visibility of $\nu=0.985$. The multi-photon emission probability values for each measurement are listed below in Sec~\ref{ssec:g2}.

\subsection{Michelson interferometry}
Michelson interferometry, i.e., the measurement of the first-order coherence $g^{(1)}(t)$, is used to determine the coherence properties of a light source and, thereby, further quantify its optical quality. 
A sketch of a Michelson interferometer is given in Fig.~\ref{sfig:setup}e).
The incoming signal is evenly divided into two interferometer arms using a 50:50 beam splitter cube (BS). Each arm contains a retroreflector (RR) mounted on a linear motorized stage (travel range \SI{30}{cm}) and a linear piezo stage (travel range \SI{2}{mm}) respectively, to redirect the signal back to the BS, where the two beams interfere. One output of the BS is coupled into a multimode fiber and sent to the spectrometer to record the intensity.
The integrated intensity of the selected emission line is recorded as a function of the piezo stage position in steps of \SI{20}{nm}. This changes the relative path length, causing a phase shift that generates an interference pattern. The resulting sinusoidal intensity profile is fitted with a cosine function. This allows us to extract the interference fringe visibility by
\begin{equation}\label{equ::vis}
    v=\frac{I_{max}-I_{min}}{I_{max}+I_{min}}.
\end{equation}
This measurement is repeated for increasing path length of the second arm (i.e. different positions of the linear stage) until the interference fringes disappear and therefore $v\approx0$.
Here, we assume that the path length difference introduced by the piezo stage is negligible in comparison to the path length difference introduced by the linear stage.
The visibility as a function of the relative delay time, i.e. the path length difference divided by the speed of light, is fitted using the Fourier-transform (FT) of a Voigt function, which is the convolution of a Gaussian and a Lorentzian. From the fit we extract an approximate linewidth \cite{Olivero1977} $\Gamma \approx 0.5346f_L+\sqrt{0.2166f_L^2+f_G^2}$, with the FWHM widths in energy domain $f_{L,G}$ of the Lorentzian and Gaussian contribution.

\section{Quantum confined Stark effect}
\begin{figure}
    \centering
    \includegraphics[]{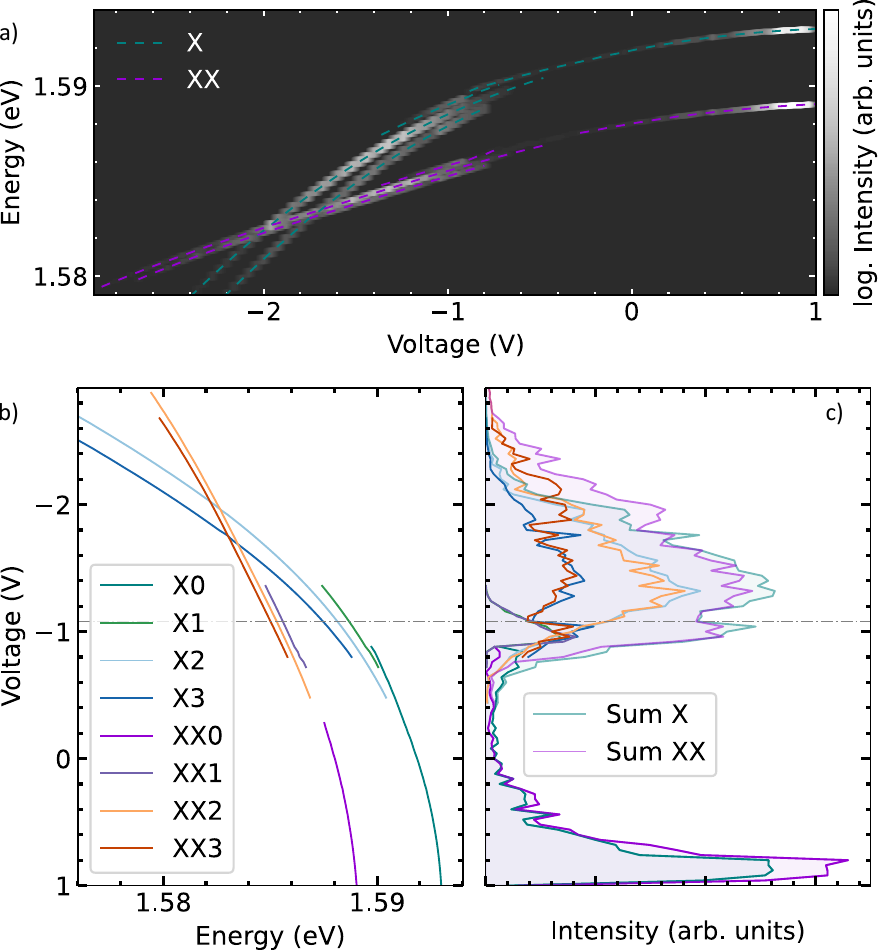}
    \caption{a) Spectra under two-photon excitation of several XX-X replicas of QD~1 as a function of the voltage. See text for the reconstruction of the plot. All emission lines are fitted using Gaussian functions. The fitted energies are plotted as dashed lines and displayed in b). Note that b) corresponds to Fig.~1~c) in the main paper. c) Intensities of the XX and X replicas as well as total XX and X intensity independent of replicas.  The gray dash-dotted line indicates \SI{-1.08}{V}, where we compared replicas 1--3 (table \ref{tab:replica_compare}). } 
    \label{sfig:tpe_replica_int}
\end{figure}

Figure~\ref{sfig:tpe_replica_int}~a) shows a 2D plot with spectra of several replicas of QD~1 under two-photon excitation at different voltages. The pump laser energy was tuned to the resonance of each replica pair. In order to present the data more clearly, only the replicas are shown, whereas residual laser and other small emission lines are cut out. This is done by showing only a few pixels around each emission line for each spectrum. Each emission line is fitted with a Gaussian function to obtain the energy. The results are shown as dashed lines as well as isolated in Fig.~\ref{sfig:tpe_replica_int}~b). 
We fitted these curves using the quadratic equation \cite{fry2000} 
\begin{equation}
   E(F) = E_0 - \alpha F - \beta F^2,
   \label{equ::qcse}
\end{equation}
describing the quantum confined Stark effect. $E_0$ is the transition energy at zero field, $\alpha$ can be interpreted as the permanent dipole of the $\ket{X}$ and $\beta$ its polarizability. Here we substituted $F= \frac{V_b-V}{D}$, with the built-in voltage $V_b=+\SI{1.7}{V}$ and the thickness of the intrinsic layer $D=\SI{305}{nm}$. 
Figure~\ref{sfig:tpe_replica_int}~c) shows the integrated intensity of the emission lines, corresponding to the XX and X replicas in Fig. \ref{sfig:tpe_replica_int}~b). The XX0 and X0 are most prominent around \SI{0.8}{V}. Towards lower voltages, the emission is quenched due to charging of the QD. Around \SI{-0.8}{V}, the replicas appear, with XX2 and X2 being the brightest for almost the entire range. The maximum of the summed up intensities of all replicas is similar to the intensity of the XX0 and X0, where the QD is blinking-free. From this, one can argue that the total emission efficiency of all the replicas together is as high as around \SI{0.8}{V}. At even higher negative voltages, additional replicas appear, which are not studied here. 

\begin{table}[!ht]
    \centering
    \setlength{\extrarowheight}{10pt}
    \begin{tabular}{l|l|l|l|l}
Replica & 0 & 1 & 2 & 3 \\ \hline \hline
$E_{0,XX}$ (eV)& 1.589018(8) & 1.5904(6) & 1.59217(8) & 1.59136(13) \\ \hline
$E_{0,\ket{XX}}$ (eV) & 3.181929(10) & 3.1811(7) & 3.17913(12) & 3.1783(2) \\ \hline
$\alpha_{XX}$ ($\frac{\mathrm{eV\cdot nm}}{\mathrm{V}}$)& -0.142(4) & 0.21(12) & 0.629(16) & 0.53(3) \\ \hline
$\alpha_{\ket{XX}}$ ($\frac{\mathrm{eV\cdot nm}}{\mathrm{V}}$) & -0.335(6) & -0.61(15) & -1.03(3) & -1.20(4) \\ \hline
$\beta_{XX}$ ($\frac{\mathrm{eV\cdot nm}^2}{\mathrm{V^2}}$) & 58.2(6) & 35(7) & 13.6(7) & 18.0(11) \\ \hline
$\beta_{\ket{XX}}$ ($\frac{\mathrm{eV\cdot nm}^2}{\mathrm{V^2}}$) & 125.2(7) & 149(8) & 181.5(11) & 202.0(19) \\ \hline
$\frac{\beta_{\ket{XX}}}{2e}$ ($\frac{\mathrm{nm}^2}{\mathrm{V}}$)& 62.6(4)&75(4)&90.8(5)&101.0(10) \\ \hline \hline
$E_{0,\ket{X}}$ (eV) & 1.592911(7) & 1.5906(4) & 1.58697(9) & 1.58700(15) \\ \hline
$\alpha_{\ket{X}}$ ($\frac{\mathrm{eV\cdot nm}}{\mathrm{V}}$) & -0.193(4) & -0.83(8) & -1.66(17) & -1.73(3) \\ \hline
$\beta_{\ket{X}}$ ($\frac{\mathrm{eV\cdot nm}^2}{\mathrm{V^2}}$) & 67.0(4) & 114(5) & 167.9(8) & 184.0(15)\\ \hline
$\frac{\beta_{\ket{X}}}{e}$ ($\frac{\mathrm{nm}^2}{\mathrm{V}}$) & 67.0(4) & 114(5) & 167.9(8) & 184.0(15)
\caption{Fit values for XX and X of each replica of QD~1 from Fig.~\ref{sfig:tpe_replica_int}}
\label{tab:beta}
    \end{tabular}
\end{table}

Table~\ref{tab:beta} shows the fit parameters of the zero field energy $E_0$, the permanent dipole $\alpha$ and the polarizability $\beta$ for XX and XX photons for the replicas 0--3 in the data shown in Fig.~\ref{sfig:tpe_replica_int}~b). In addition to the values obtained for the XX and X lines, it is interesting to evaluate also the values of $E_0$, $\alpha$ and $\beta$ for the excitonic states giving rise to these emission lines. For the $\ket{X}$, these directly correspond to the fit parameters. For the $\ket{XX}$, the respective fit parameters of both photons have to be summed up: $E_{0,\ket{XX}}=E_{0,XX}+E_{0,X}$, $\alpha_{\ket{XX}}=\alpha_{XX}+\alpha_{X}$ and $\beta_{\ket{XX}}=\beta_{XX}+\beta_{X}$.
Although we used a quadratic fit function, we expect the real function for $E(F)$ to be slightly asymmetric, as the QD potential is asymmetric in the growth direction.

\subsection{Polarizability}
Since the $\ket{XX}$ comprises the double number of charges compared to the $\ket{X}$, we introduce $\frac{\beta}{q}\equiv\frac{\partial d}{\partial F}$ 
to better compare both states. This quantity describes the change of the average distance $d$ of the charges in the dipole under electric field changes. We use this to characterize the change in wave function overlap. For the $\ket{X}$, the charge of the dipole is $q=e$, whereas it is $q=2e$ for the $\ket{XX}$.
As seen in table~\ref{tab:beta} $\frac{\beta}{q}$ is higher for the $\ket{X}$ than for $\ket{XX}$ for all replicas.
Whereas for positive voltages (replica 0), the values are similar, the difference increases with increasing field and $\frac{\beta_{\ket{X}}}{e}$ becomes more than $1.5\times$ higher than $\frac{\beta_{\ket{XX}}}{2e}$. 
This is visible in Fig.~\ref{sfig:tpe_replica_int}~b), as the XX energies show much lower curvatures compared to the X energies. 

\subsection{Wave function overlap}
The lower polarizability of the $\ket{XX}$ can be qualitatively interpreted as a partial screening of the external electric field. Correlation effects, which are more pronounced for the $\ket{XX}$ compared to the $\ket{X}$, will tend to minimize Coulomb repulsion among carriers of the same sign and maximize electron-hole attraction for any given external field, thus reducing the effect of the external perturbation. 
This leads to a higher sensitivity of the $\ket{X}$ to changes in the electric field. In addition to the higher tunability of the X emission energy, higher fields also lead to a lower wave function overlap compared to the $\ket{XX}$. This results in tunability of the $\ket{X}$ lifetime by a factor of $3.5$. Furthermore, the $\ket{X}$ also shows a higher sensitivity to charge noise, leading to increased linewidth broadening. For the $\ket{XX}$, on the other hand, the lifetime increases less and the linewidth remains transform limited.

\subsection{Permanent dipole}
The shortest $\ket{X}$ lifetime we measured is $\approx\SI{175}{ps}$ at an applied voltage of $V\simeq\SI{1}{V}$. Unfortunately, we cannot access the neutral X in the zero electric field region ($V=V_b\simeq\SI{1.7}{V}$), since the QD emission is dominated by highly negatively-charged excitonic states in this regime. For comparison, GaAs QDs in non-diode samples exhibit $\ket{X}$ lifetimes of $\approx 200-\SI{250}{ps}$ \cite{DaSilva2021}. This suggests that the electron-hole wave function overlap is not maximal in non-diode samples (zero external electric field). Specifically, from the quadratic fit of the data for QDs measured in this work, we find that the permanent electric dipole at zero field is negative (see negative values of $\alpha_X$ in Table~\ref{tab:beta}), indicating that the barycenter of the hole wave function is located below the electron in growth direction. Interestingly, this result is different from that obtained in Ref.~\cite{Huang2021a}, where a dipole with opposite sign was observed, possibly due to strain-induced effects due to bonding on a different substrate in the QDs studied there or to differences in the structural properties of the QDs. 
A permanent dipole has previously been observed in several works on InGaAs QDs \cite{fry2000, barker2000,jin2004, Finley2004, mar2017} and its sign manipulated by strain~\cite{Aberl2017}. 
With increasing fields ($V\approx1-\SI{0.5}{V}$), the charge carriers are first pushed closer together (increasing the wave function overlap and thus shortening the lifetime) until they swap positions and the electron is located in the tip of the nanohole. Now, with increasing fields, the charge carriers are pulled apart, leading to decreasing wave function overlap and therefore increasing lifetime.

\section{Replicas of the XX-X pairs}

\subsection{Origin of the replicas}
\begin{figure}
    \centering
    \includegraphics[width = \textwidth]{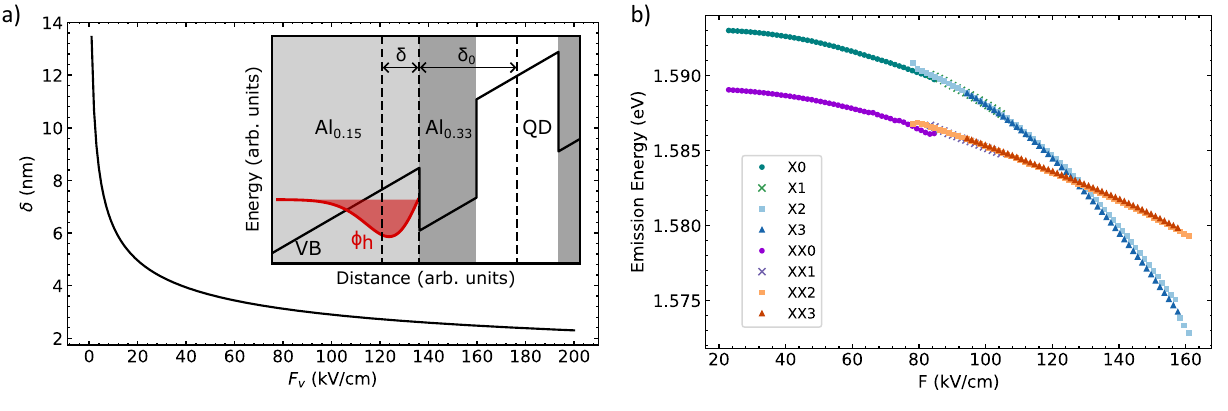}
    \caption{Inset: Zoom in from Fig.~1~b) in the main manuscript: the valence band edge forms a triangular potential below the QD, where holes can be trapped. $\delta_0=\SI{11.5}{nm}$ is the distance between the Al$_{0.33}$Ga$_{0.67}$As/Al$_{0.15}$Ga$_{0.85}$As interface and the center of the QD. $\delta$ is the field-dependent distance of the trapped hole to the interface. a) Calculated distance $\delta$ between the barycenter of the hole wave function in the triangular potential well and the Al$_{0.33}$Ga$_{0.67}$As layer, indicated in the inset as a function of the electric field $F_v$ given by the applied voltage. b) Emission energy of the QD~1 replicas as a function of the total electric field $F$ resulting from the applied voltage, and the field created by one, two or three positive charges (and their mirror image at the n-doped layer) at distance $\delta_0+\delta$.} 
    \label{sfig:shift}
\end{figure}

Since we observe discrete replicas in our spectra, we make discrete changes in the electrical field responsible, which we attribute to discrete numbers of charges in the vicinity of the QD. 
All investigated QDs show red shifted replicas (see Fig.~\ref{sfig:tpe_replica_int}~a,b)), indicating that the additional charge carriers enhance the applied field. Consequently, we expect either electrons above the quantum dot (toward the p-contact) or holes below the quantum dot (on the n-side of the diode). Since the band structure lacks a potential structure to trap electrons, we infer that holes are likely trapped in the triangular-shaped potential within the Al$_{0.15}$Ga$_{0.85}$As layer (see Fig.~\ref{sfig:shift}~a, inset))~\cite{Houel2012}.
In order to find out whether these holes could lead to the observed shift in energy, we first examine the situation of a point charge located right below the QD and reduce the situation to a one-dimensional problem. In this scenario, we calculate the field with the contribution of the applied voltage $F_v=\frac{V_b-V}{D}$, the contribution from a point charge $F_c$ at distance $\delta+\delta_0$ from the center of the QD and its mirror charge in the n doped layer $F_m$. As indicated in the inset of Fig.~\ref{sfig:shift}~a), $\delta_0=\SI{11.5}{nm}$ is the distance between the center of the QD and the interface between the Al$_{0.15}$Ga$_{0.85}$As and Al$_{0.33}$Ga$_{0.67}$As layers. $\delta$ is the distance between the barycenter of the trapped additional hole wave function and the interface. Since the triangular well changes with the applied field, $\delta$ is also voltage dependent. This is calculated by solving the Schrödinger equation for a triangular well resulting in Airy functions, as plotted in Fig.~\ref{sfig:shift}~a). $\delta$ is determined by the expectation value. From this, the resulting electric field $F$ is calculated to be:
\begin{equation}
    F = F_v+n\frac{1}{(\delta(F_v)+\delta_0)^2\cdot4\pi \epsilon_0\epsilon}+F_m
\end{equation}
with $n$ being the number of holes (1 for X1/XX1, 2 for X2/XX2, 3 for X3/XX3), $\epsilon_0$ being the vacuum permittivity and $\epsilon$ being the permittivity in Al$_{0.33}$Ga$_{0.67}$As at \SI{4}{K}, extrapolated from literature \cite{adachi1985, strzalkowski1976}. 
Figure~\ref{sfig:shift}~b) shows the emission energies of QD~1 as a function of the corrected field $F$. It shows that the XX and X emission lines of all replicas now nicely follow the same behavior. We thus conclude that the discrete jumps observed in the spectra are consistent with the presence of a number of $n$ holes at the interface between the two AlGaAs layers, right below the QD. 

What remains to be explained is how holes are generated and why they seem to get trapped right below the QD rather than at an arbitrary location in the plane of the interface.

\subsection{Photogeneration of holes near QDs}
\begin{figure}
    \centering
    \includegraphics[width = 3.3in]{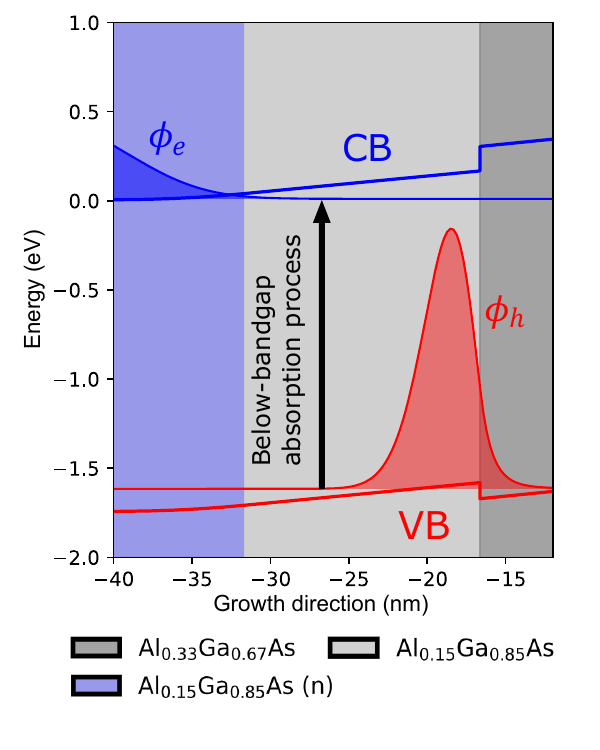}
    \caption{Schematics of the absorption of laser photons via the Franz-Keldysh effect. After absorption, electrons and holes are separated by the electric field with electrons drifting to the n-doped layer and holes getting trapped at the Al$_{0.33}$Ga$_{0.67}$As/Al$_{0.15}$Ga$_{0.85}$As interface}
    \label{sfig:Franz_Keldysh_effect}
\end{figure}

As discussed above, we expect holes being confined at the triangular potential due to the band bending in the Al$_{0.15}$Ga$_{0.85}$As layer. The explanation of the presence of these holes is not trivial, since (i) the electric field leads to full depletion of holes close to the n-region of the diode; (ii) the laser energy ($E_{\text{ph}} \approx \SI{1.59}{eV}$) is smaller than the band gap ($E_\text{g} \approx \SI{1.73}{eV}$), prohibiting direct optical excitation of electron-hole pairs; and (iii) possible tunneling of holes from the QD to the triangular potential is blocked by the large tilt of the bands at high electric fields.
Instead, we ascribe the generation of these holes to the absorption of laser photons in the undoped Al$_{0.15}$Ga$_{0.85}$As layer via the Franz-Keldysh effect, which describes the appearance of an absorption tail below the bandgap energy at high electric fields. \cite{chuang2009physics}
Specifically, a laser photon can generate an electron-hole pair, as illustrated in Fig.~\ref{sfig:Franz_Keldysh_effect}. The electron wave function is extended over the n-type region with an exponential tail leaking into the Al$_{0.15}$Ga$_{0.85}$As layer and the hole is trapped at the triangular potential well.
The finite extension of the Al$_{0.15}$Ga$_{0.85}$As barrier (thickness: $d=\SI{15.1}{nm}$) does not allow for the absorption of photons at too low electric fields. Instead, we expect an onset of the Franz-Keldysh, when the electric field $F$ in the diode is high enough so that the potential difference $e\cdot F \cdot d$ across the Al$_{0.15}$Ga$_{0.85}$As layer compensates the energy difference $E_\text{g} - E_\text{ph} \approx \SI{0.14}{eV}$. In our diode structure, this condition is fulfilled for a voltage of $V \approx \SI{-1.16}{V}$. Starting from here, the absorption coefficient of the layer increases with decreasing voltage. Indeed, we see the maximum intensity of the replica lines in the voltage range of about \SIrange{-1.0}{-1.8}{V}, as visible in Fig.~\ref{sfig:tpe_replica_int}~c). We expect these holes to be long-lived compared to the excitonic lifetimes, since electrons from the n-doped layer are pushed away from the interface by the large electric field. 

\subsection{Localization of the holes close to the QD}
While the confinement of the holes in the triangular well along the growth direction is straightforward, this only serves to trap them in plane. The fact that we observe discrete emissions from the replicas, together with the energy shift observed, leads us to conclude that the holes are trapped directly below (in growth direction) the QD. This requires a lateral confinement potential, and we explored several possible explanations: (i) the electric dipole of a polarized $\ket{X}$ or $\ket{XX}$ in the QD, producing an attractive potential on holes located at the interface; (ii) the local strain caused by the QD itself due to the small lattice mismatch between GaAs and surrounding barrier; (iii) local interface roughness. Both contributions (i) and (ii) induce an attractive potential well below \SI{1}{meV}, so that the discrete jumps in emission should disappear at temperatures well above \SI{1}{meV}/$k_B\simeq\SI{12}{K}$.  To test these ideas, we examined the emission at \SI{30}{K} to see if the holes would be thermally excited and the replicas would disappear. Since this was not the case, these contributions can not solely explain the lateral localizations of the holes.

Concerning (iii) we note that the rate of formation of Al droplets in the LDE method increases with the roughness of the substrate\cite{Kruck2024}. We speculate that an area of increased roughness at the interface between the 15\% and 33\% Al layers might transfer to the overlying layer for LDE growth and favor the formation of droplets there. This could lead to trapping of holes, specifically directly below the QD.

\subsection{Replica intensities}
Figure~\ref{sfig:tpe_replica_int}~c) shows the integrated intensity of the XX and X replicas separated, as well as the sum of all XX and X emission lines.
For almost all voltages, the XX intensity is roughly equal to the X intensity for each replica. From $V=\SI{0}{V}$ to $V=\SI{-0.7}{V}$ the overall intensity is very low, since positively charged states are dominant in this region. For lower voltages, replicas 1--3 start to appear, while the initial replica 0 disappears. Replica 2 is predominant for the largest voltage range, namely from $V=\SIrange{-1.08}{-1.7}{V}$. It is therefore used for almost all measurement points in the main paper. Only at $V=\SI{-2.04}{V}$ replica 3 is chosen, as it is brighter. After the crossing of the XX and X emission energies, the intensity of the X decreased faster than the XX. This, together with the appearance of even more replicas, limits the maximum tuning range. 

\subsection{Comparison of replicas at the same voltage}
In order to prove that replicas behave similarly, we compare the three replicas (replicas 1-–3) observed at an applied voltage of \SI{-1.08}{V}, where all replicas show a reasonable brightness. The summarized results in Tab.~\ref{tab:replica_compare} show that all replicas exhibit similar behavior. We therefore focus on the most prominent replica for each voltage in the main paper, mostly replica 2 (see. Fig.~\ref{sfig:tpe_replica_int}~c) ). Only at the last data point at $\SI{-2.04}{V}$ replica 3 dominated.
These findings and the energy shift behavior shown in Fig.~\ref{sfig:shift} makes us confident that replicas are not necessary to tune the ratio between $\ket{XX}$ and $\ket{X}$ lifetime and that a device featuring no trap for holes would still be suitable to this aim.

\begin{table}[!ht]
    \centering
    \begin{tabular}{l|l|l|l}
        Replica & 1 & 2 & 3 \\ \hline \hline
        XX $\tau$ (ps) & 119(4) & 133(3) & 122(3) \\ \hline
        X $\tau$ (ps) & 227(7) & 227(4) & 252(3) \\ \hline
        $r$ & 0.52(3) & 0.59(2) & 0.484(13) \\ \hline
        XX $\Gamma$ ($\mu$eV)&10(2)&9.4(12)&7.7(2)\\ \hline
        XX $\frac{\Gamma}{\Gamma_0}$&1.2(3)&1.2(2)&1.0(4)\\ \hline
        X $\Gamma$ ($\mu$eV)&4.9(6)&4.6(5)&5.6(4) \\ \hline
        X $\frac{\Gamma}{\Gamma_0}$&1.7(3)&1.6(2)&2.1(2)\\ \hline
        XX $g^{(2)}(0)$ & 0.0035(10) & 0.0069(10) & 0.0120(5) \\ \hline
        X $g^{(2)}(0)$ & 0.006(6) & 0.0070(5) & 0.008(1) \\ \hline
        raw XX $V_{HOM}$ & 0.607(10) & 0.617(12) & 0.568(14) \\ \hline
        corr XX $V_{HOM}$ & 0.625(10) & 0.638(12) & 0.594(15) \\ \hline
        raw X $V_{HOM}$ & 0.535(12) & 0.536(10) & 0.501(12) \\ \hline
        corr X $V_{HOM}$ &0.553(12) & 0.555(12) & 0.519(13) 
    \caption{Lifetimes, lifetime ratio $r$, linewidths $\Gamma$, also with respect to their transform limited linewidth $\Gamma_0$, $g^{(2)}(0)$, and HOM visibilities of the XX and X for the replicas 1-3, which are present at a voltage of \SI{-1.08}{V}.}
    \label{tab:replica_compare}
\end{tabular}
\end{table}
\newpage

\section{Additional Data}

\subsection{Indistinguishability of the resonantly excited negative trion}

\begin{figure}
    \centering
    \includegraphics[width = 6.6in]{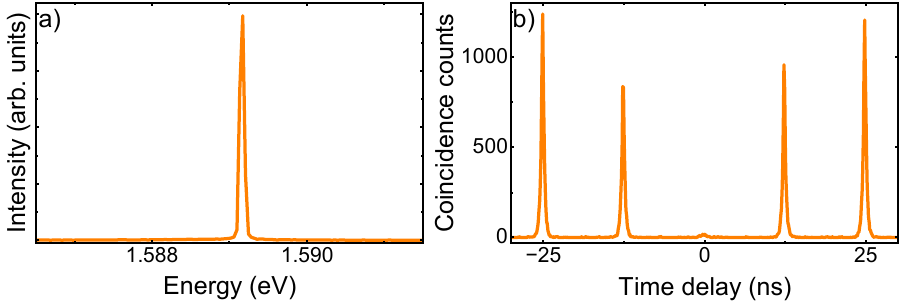}
    \caption{a) Spectrum of the negative trion of QD~1 at a voltage of \SI{1.03}{V} under resonant fluorescence conditions. b) Histogram of the corresponding HOM measurement.}
    \label{sfig:T}
\end{figure}
To assess whether our sample suffers from noise sources or the HOM setup is not well aligned, we measure the HOM interference visibility of photons emitted by a negative trion under resonance fluorescence conditions. 
The unfiltered spectrum of this transition is shown in Fig.~\ref{sfig:T}~a) and the corresponding coincidence histogram is shown in Fig.~\ref{sfig:T}~b). The measured raw HOM visibility is 0.944(4). Correction for a small multi-photon contribution ($g^{(2)}(0)=0.009(2)$) and imperfect beam overlap at the beam splitter yield a corrected HOM visibility of 0.991(6). 

\subsection{Single-photon purity}
\label{ssec:g2}
\begin{figure}
    \centering
    \includegraphics[width = 3.3in]{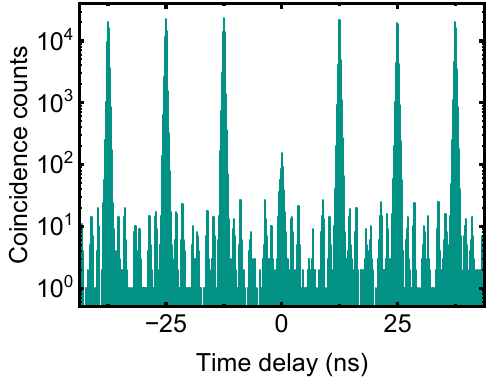}
    \caption{Second-order autocorrelation histogram of the XX at an applied voltage of \SI{-1.08}{V} yielding $g^{(2)}(0)=0.0035(10)$.}
    \label{sfig:g2}
\end{figure}

We measure the second-order autocorrelation function for the XX and X, as described in Sec.~\ref{subsec:autocorrelation}, to verify good filtering of the emission line of interest and for the possibility to correct the measured HOM visibilities for remaining multi-photon contributions. The analyzed $g^{(2)}(0)$ values are listed in Table~\ref{tab: g2}. An example measurement is shown in Fig.~\ref{sfig:g2}.

\begin{table}[]
    \centering
    \begin{tabular}{c|c|l|l}
      Voltage (V)&Replica&XX $g^{(2)}(0)$   &X $g^{(2)}(0)$ \\ \hline \hline 
      0.9& 0& 0.008(1)&0.015(2)\\ \hline
      0.8&0&0.006(1)&0.004(1)\\ \hline
      0.7&0&0.012(6)&0.0070(3)\\ \hline
      0.6 & 0&0.010(7) &0.010(1)\\ \hline
      -1.08&1&0.0035(10)&0.006(6)\\ \hline
      -1.08&2&0.0069(10)&0.0070(5)\\ \hline
      -1.08&3&0.0120(5)&0.008(1)\\ \hline
      -1.4&2&0.0151(12)&0.0106(11) \\ \hline
      -1.6&2&0.0180(16)&0.0105(6) \\ \hline
      -1.7&2&0.039(4)&0.1140(14)\\ \hline
      -2.04&3&0.0187(17)&
       
    \end{tabular}
    \caption{Single-photon purity $g^{(2)}(0)$ of the XX and X for different voltages. }
    \label{tab: g2}
\end{table}

\subsection{Data on additional quantum dots}
All other QDs consistently show replicas, however, with different replicas being more or less pronounced. In the following, we present the data of QD~2, included in Fig.~3~d) in the main manuscript, as well as replicas of QD~3.

\subsubsection{Data of QD~2}
\begin{table}[!ht]
    \centering
    \begin{tabular}{l|l|l|l|l|l}
        Voltage (V) & 0.85 & -1.55 & -1.8 & -2.1 & -2.2 \\ \hline
        Replica & 1 & 3 & 3 & 4 & 4 \\ \hline \hline
        XX $\tau$ (ps) & 104(4) & 138(3) & 154(3) & 168(4) & 155(5) \\ \hline
        X $\tau$ (ps) & 175(4) & 287(5) & 351(8) & 536(30) & 569(26) \\ \hline
        r & 0.59(4) & 0.27(2) & 0.31(3) & 0.44(2) & 0.48(2) \\ \hline
        XX $g^{(2)}(0)$ & 0.007(1) & 0.010(1) & 0.019(2) & 0.006(1) & 0.006(1) \\ \hline
        X $g^{(2)}(0)$ & 0.005(1) & ~ & ~ & ~ & ~ \\ \hline
        raw XX $V_{HOM}$ & 0.608(30) & 0.655(17) & 0.662(12) & 0.701(11) & 0.694(12) \\ \hline
        corr XX $V_{HOM}$ & 0.636(31) & 0.688(17) & 0.708(13) & 0.733(11) & 0.724(12) \\ \hline
        raw X $V_{HOM}$ & 0.581(18) & ~ & ~ & ~ & ~ \\\hline 
        corr X $V_{HOM}$ & 0.605(19) & ~ & ~ & ~ & ~ 
    \end{tabular}
\end{table}

\subsubsection{Replicas of QD~3}
\begin{figure}
    \centering
    \includegraphics[]{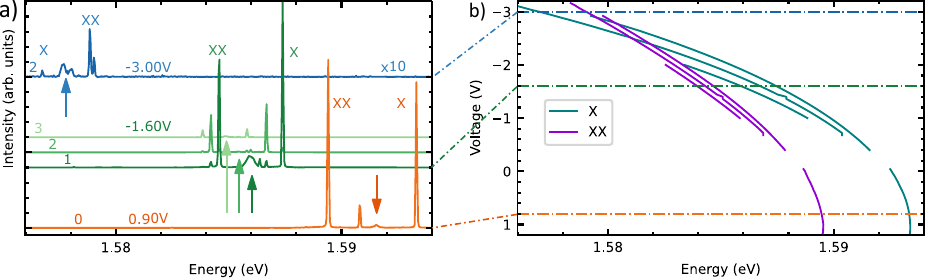}
    \caption{a) Spectra of QD~3 at three different voltages under $\pi$-pulse two-photon excitation. Depending on the voltage, several XX-X replicas (labeled 0--3) could be observed, sometimes simultaneously. For each spectrum, the laser energy was adjusted to match the two-photon resonance. Small contributions from other replicas come from phonon-assisted excitation. The wavelength of the excitation laser is indicated with arrows. b) fitted energies of the XX and X photons as a function of the applied voltage are shown. Each data point is extracted from a spectrum under two-photon excitation of the respective replica.} 
    \label{sfig:qd2_replica}
\end{figure}

\begin{figure}
    \centering
    \includegraphics[width=0.5\linewidth]{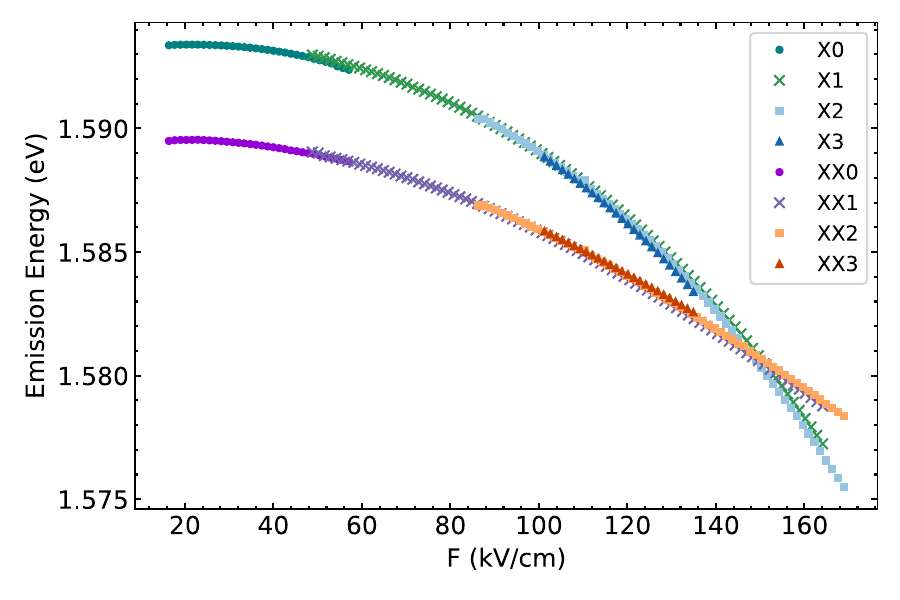}
    \caption{Emission energy as a function of the total electric field $F$ for QD~3. The electric field contains the contribution of one, two or three positive charges located at a distance $\delta_0+\delta$ from the QD center, where $\delta$ is the distance of the hole to the interface below the QD layer. The $\delta_0$ is set to \SI{11.5}{nm}, which is roughly the distance from the interface between the Al$_{0.15}$Ga$_{0.85}$As and the Al$_{0.33}$Ga$_{0.67}$As layers to the center of the QD.}
    \label{sfig:shiftQD2}
\end{figure}

In Fig.~\ref{sfig:qd2_replica} we show a graph similar to Fig.~1~c) and d) from the main text but for QD~3. Recording and reconstruction of the graphs was explained above and in the main manuscript.Figure~\ref{sfig:qd2_replica}~a) shows spectra of every replica pair under TPE for a certain voltage, whereas b) shows the energy shift of each emission line as a function of the voltage.  Compared to QD~1 of the main paper, the first replica is much more pronounced, while especially the third replica is only visible in a narrow voltage range. For $V=\SI{0.90}{V}$ (orange) only replica 0 is visible. For $V=\SI{-1.60}{V}$ (green) and $V=\SI{-3.00}{V}$ (blue), three (replica 1-3) and one (replica 2, 10$\times$ enhanced in brightness) replicas are visible. 
Figure~\ref{sfig:shiftQD2} shows that the energy shifts of the replicas can also be explained by the field contribution from additional holes, as discussed above.

\bibliography{supportingReferences}